\documentclass[aps,prl,twocolumn,showpacs]{revtex4} 
\usepackage{amsmath} 
\usepackage{amssymb} 
\usepackage{graphicx} 
\usepackage{graphics}
\usepackage{epsfig}
\usepackage{slashed}
\newcommand\ba{\begin{eqnarray}}
\newcommand\ea{\end{eqnarray}}
\newcommand\alb{\begin{align}}
\newcommand\ale{\end{align}}
\newcommand\be{\begin{equation}}
\newcommand\ee{\end{equation}}

\usepackage{amsmath} 
\usepackage{amssymb}

\begin{document}

		\title{Periodic interference structures in the time-like
	proton form factor}


		\author{Andrea Bianconi} 
    \affiliation{\it Dipartimento di
	Ingegneria dell\!~$^\prime$Informazione, Universit\`a degli Studi di Brescia
	and Istituto Nazionale di Fisica Nucleare, Gruppo Collegato di
	Brescia, I-25133, Brescia, Italy}

		\author{Egle~Tomasi-Gustafsson} \affiliation{\it CEA,IRFU,SPhN,
	Saclay, 91191 Gif-sur-Yvette Cedex, France}

		\date{\today}

		\begin{abstract}
An intriguing and elusive feature of the time-like hadron 
form factor is the possible presence of an imaginary part associated to 
rescattering processes. We find evidence of that in  the recent and 
precise data on the proton time-like 
form factor measured by the BABAR collaboration. By plotting these 
data as a function of the 3-momentum of 
the relative motion of the final proton and antiproton, 
a systematic sinusoidal modulation is highlighted in the 
near-threshold region. 
Our analysis attributes this pattern  
to rescattering processes at a relative distance of 0.7-1.5 fm between 
the centers of the forming hadrons. This distance implies a large  
fraction 
of inelastic processes in $\bar{p}p$ interactions, 
and a large imaginary part in the related $e^+e^- \rightarrow \bar{p}p$ 
reaction because of unitarity. 

		\end{abstract}
\maketitle


Electromagnetic hadron form factors (FFs) are 
fundamental quantities which describe the internal structure of the hadron  
(for a recent review see Ref. \cite{Pacetti:2015iqa}). 
FFs enter explicitly in the coupling of a virtual photon with the hadron 
electromagnetic current, and can be directly compared to hadron 
models which describe dynamical properties of hadrons. They are 
experimentally accessible through the knowledge of the differential cross 
section and the polarization observables. 

Traditionally most data on FFs come from electron-proton elastic 
scattering. It is assumed that the interaction occurs through the 
exchange of a single virtual photon which carries a four  
momentum transfer squared $q^2$. In this kinematical region (space-like, SL) 
the virtual 
photon-proton coupling and the electron-proton scattering cross 
section are described via two real FFs, electric, $G_E$, and magnetic, $G_M$. 

FFs have been also studied in the time-like (TL) region of momentum 
transfer squared. They are measured through the reactions
\ba
&e^+\ +\ e^-\ &\to\ \bar p\ +p\  ,
\label{eq:eq1}\\
& \bar p\ +\ p\ &\to\  e^+\ +\ e^- \ ,
\label{eq:eq2}
\ea
where a hadron pair is formed by or annihilated into a virtual 
photon. In the following we will 
refer to the former reaction, when not otherwise specified. 

Assuming single photon exchange, the unpolarized cross section
contains the squared moduli of two TL FFs (electric and magnetic FF), 
which are complex-valued functions of $q^2$. The imaginary part is expected to be large and information on the relative phase between $G_E$ and $G_M$ can be extracted from 
single spin polarization experiments \cite{Dubnickova:1992ii} which are presently out of reach. In this letter evidence for periodic structures in TL FF data is reported and related to their complex nature.

Many models for the hadron coupling to the virtual photon have been 
developed and applied to the calculation of SL FFs. 
Some of them may be analytically 
continued to the TL region. This is the case for approaches based on 
vector meson dominance  \cite{Bijker:2004yu, Adamuscin:2005aq} and 
dispersion relations  \cite{Belushkin:2006qa,Lomon:2012pn}.  
Constituent quark models in light front dynamics 
may be applied  \cite{deMelo:2003uk}, as well as approaches based on 
AdS/QCD correspondence \cite{Brodsky:2007hb}. A 
phenomenological picture has been recently proposed for an interpretation of FFs in both the SL 
and TL regions \cite{Kuraev:2011vq}. 
	
The individual determination of the electric and magnetic TLFF is obtained in principle, from 
the angular distribution of reactions (\ref{eq:eq1},\ref{eq:eq2}), but until now the luminosity was not sufficient. 
The various experimental results are therefore compared on the 
basis of a generalized FF\cite{Bardin:1994am}, which is related to the 
unpolarized cross section by:
$\sigma$:
\be
|F_p|^2=\displaystyle\frac{3\beta q^2 \sigma}
{2\pi\alpha^2 \left(2+\displaystyle\frac{1}{\tau}\right)}, 
\label{eq:Fp}
\ee
where $\alpha=e^2/(4\pi)$, $\beta=\sqrt{1-1/\tau}$, 
$~\tau={q^2}/(4M^2)$, and $M$ is the proton mass. 

Even in these simplified terms, 
it has long been difficult to analyze with precision the behavior of the 
data over a broad kinematic range because of the uncertainties and of 
the matching of data from different experiments which covered limited  
$q^2$ regions. The recent data by the BABAR  collaboration 
\cite{Lees:2013xe,Lees:2013uta} cover with reasonable 
continuity a region ranging from slightly over the $\bar{p}p$ threshold 
to $q^2$ $\approx$ 36 GeV$^2$. In particular about 30 data points have been extracted  
in the region $q^2$ $<$ 10 GeV$^2$, with a relative  
error lower than 10 \%. These features allow for a refined analysis of 
the systematic behavior of the TLFF, where 
large-scale and small-scale (in $q^2$ sense) properties of the data 
distribution may be scrutinized. 


From now on, we use the expression "near-threshold region" to 
indicate a $q^2$-range extending from the threshold of 
the $\bar{p}p$ channel  up to $q^2$ $\approx$ 9 GeV$^2$ 
(with the convention $c=\hbar=1$). 
In this kinematic region, two different scales participate: 
the total energy  of the colliding $e^+e^-$ pair is $>2M\approx 1.9$ GeV, while the kinetic energy of the created $\bar p p$ 
pair is relatively small. Therefore one may expect to observe complex 
effects where a highly relativistic formation picture expressed in 
terms of quarks and gluons coexists with non relativistic interactions 
of two slow hadrons leaving the formation zone. 

Proton-antiproton interactions in the near-threshold region have been studied 
in experiments at 
LEAR (see \cite{Zenoni:1999su,Zenoni:1999st} 
and 
references therein for previous data) and more recently 
at AD \cite{Bianconi:2011zz}. These measurements could not separate 
spin channels and, 
as in the case of the single effective FF of 
Eq. (\ref{eq:Fp}), these data have been mostly analyzed in terms of a single 
effective scattering amplitude, as if proton and antiproton were 
scalar particles. 

We define "large inelasticity"  when,  writing the amplitude as a sum of partial waves, at least half of the incoming 
flux is absorbed for all the partial waves of angular momentum $L$ satisfying 
$L \leq R p $, $R\simeq 1 $ fm.  The unitarity limit is reached when there is total  absorption for these waves. For the inelastic cross section $\bar p + p \to X \neq \bar p  + p$ "large inelasticity''  occurs in all the 
kinematical range of interest here, with the possible exception of the 
region 
$p$ $<$ 50 MeV \cite{Bianconi:2000ap,Bruckner:1989ew}. For $p\gg$ 100 MeV the inelastic cross section 
of $\simeq$ 40 mb is close to the black disk limit $\approx$ 50 mb. 

As a consequence, unitarity leads to a large imaginary part in the amplitude of $\bar{p}+ p\rightarrow $ {\it exclusive final states}, including (\ref{eq:eq2}).  Near the threshold 
this is rigorously stated by Watson's 
final state theorem \cite{Watson:1952ji} applied to reaction (\ref{eq:eq1}).
More in general the presence of a large transition amplitude  
$\bar{p}+p \rightarrow A_n$, where 
$A_n$ is an on-shell channel, implies a contribution 
to the imaginary part of the amplitude of (\ref{eq:eq2}) from a 
Cutkosky cut applied to the intermediate state 
of the 2-step process 
$\bar{p}+ p \rightarrow A_n \rightarrow e^+ + e^-$. 

Understanding at which extent phenomenological $\bar{p}p$ 
interactions could be final state interactions of the reaction (\ref{eq:eq1}) 
and seriously affect its amplitude,  
is however complicated by an evident "mismatch" 
of two known features of these processes: 

\noindent 1) The analysis of annihilation and scattering data has shown that colliding proton and 
antiproton do not overlap at small kinetic energies. As soon as 
these  particles come close within 1 fm of each other, they either 
annihilate or scatter in a hard-core way \cite{Bianconi:2000ap,Batty:2000vr}. 
So the wave function describing a  $\bar{p}p$ state presents a "hole" 
of size 1 fm. Within  a  momentum range of 400 MeV over the $\bar{p}p$ 
threshold, this 
property is demonstrated by counter-intuitive phenomena as the equality of 
$\bar{p}p$, $\bar{p}$D, and $\bar{p}^4$He annihilation cross sections 
at small energies \cite{Bianconi:2000ap}), and the suppressed effect 
of the electric charge in $\bar{p}$-nucleus 
annihilation 
with the paradoxical effect of $\bar{n}$ cross sections on heavy 
nuclei exceeding $\bar{p}$ ones \cite{Friedman:2014vva}. 

\noindent 2)  In the exclusive reaction (\ref{eq:eq1}) 
quark-counting rules \cite{Matveev:1973uz,Brodsky:1973kr} predict 
that in the initial stages of the formation process quarks and antiquarks 
are concentrated in a region of size $1/\sqrt{q^2}$, which means at a relative 
distance of 0.1 fm when  $q^2 \approx$ 4 GeV $^2$. 
So, in the initial formation stages the hadrons lie at a distance 
that is not normally tested in phenomenological $\bar{p}p$ interactions, 
making rescattering effects largely unpredictable. 

In order to search for signals of final state effects at small 
kinetic energies  in the data, it is more convenient to introduce variables directly 
related to the relative motion of the hadron pair. In the following 
we will use the 3-momentum $p$ of one of the two hadrons in the 
frame where the other one is at 
rest: 
\ba
p\ \equiv\ \sqrt{E^2-M^2},\hspace{0.3truecm} E\ \equiv\ q^2/(2M)\ -\ M. 
\label{eq:plab}
\ea
The 
usefulness of this variable presumes that the process can be divided into two 
stages: formation and rescattering, where the latter 
involves energies on a smaller scale than the former. This means 
that the amplitude for the process is the sum of a 
leading term due to a "bare formation" process 
taking place on a time scale $1/\sqrt{q^2}$, and a relatively 
small perturbation associated to rescattering processes taking place on a larger time scale. 

A consequence of this assumption is that the measured FFs 
can be fitted by a function of the 
form :
\begin{align}
&F(p)\ \equiv\ F_0(p)\ +\ F_{osc}(p), 
\label{eq:diff}
\\
&|F_{osc}(p)|\ \ll \ |F_0(p)|, 
\label{eq:diff2}
\\
&<F_{osc}(p)>_{\Delta p}\ \rightarrow\ 0\ for\ \Delta p\ \gg \ 1\  \mbox{GeV}^2, 
\label{eq:diff3}
\end{align}
where: 
\begin{itemize}
\item $F_0(p)$ is the translation in terms of the variable $p$ of a 
known parametrization that has been adjusted on the data in the full 
range $4M^2< q^2 \lesssim$ 36 GeV$^2$ (see below) ignoring small-scale oscillations. $F_0(p)$ is regular and smooth on the GeV/c scale. We name it 
``regular background fit''. 
\item $F_{osc}(p)$ reproduces GeV-scale or sub-GeV-scale irregularities in the 
lower part of the $p$ range. We name it ``oscillation fit''.  $<F_{osc}(p)>_{\Delta p}$ is the local average of $F_{osc}(p)$ over 
the momentum range $[p- \Delta p /2, p + \Delta p / 2]$. 
\end{itemize}
The data by the BABAR  
collaboration \cite{Lees:2013xe,Lees:2013uta} are selected for this study,  
since they are the most precise data in the near-threshold region and they cover with continuity a very large kinematic range. 
Both properties are necessary for our analysis. The  fitting 
procedure is the one of the Minuit package of root.cern.ch \cite{Brun:1997pa}, 
based on the minimization of  
$\chi^2= \sum_i (f(x_i,p_i)-y_i)^2/\sigma_i^2$, where $\sigma_i$ is 
the error on the point of coordinates $(x_i,y_i)$ and $f$ is the fitting function depending on the parameters $p_i$. The error on $p_i$ is the  interval where $\chi^2/n.d.f.$ increases by one unit ($n.d.f.$ is  the number of degrees of freedom: number of points minus number of parameters). 


To satisfy Eq.  (\ref{eq:diff3}) 
a good choice of the regular background term $F_0(p)$ is needed. 
The generalized FF has been consistently extracted at $e+e^-$ 
colliders and antiproton 
facilities. It shows a decreasing behavior as function of $q^2$, 
which was generally fitted in the experimental papers before 
the year 2006 with the function \cite{Ambrogiani:1999bh,Lepage:1979za}: 
\be
|F_{scaling}(q^2)|=\displaystyle\frac{\cal A}{(q^2)^2\log^2(q^2/\Lambda^2)}.
\label{eq:qcd}
\ee  
A good fit of the data prior to BABAR last results     
was obtained with ${\cal A}=40$ GeV$^{-4}$ and $\Lambda=.45$ GeV$^{2}$. 

\begin{figure}
\begin{center}
\includegraphics[width=8.5cm]{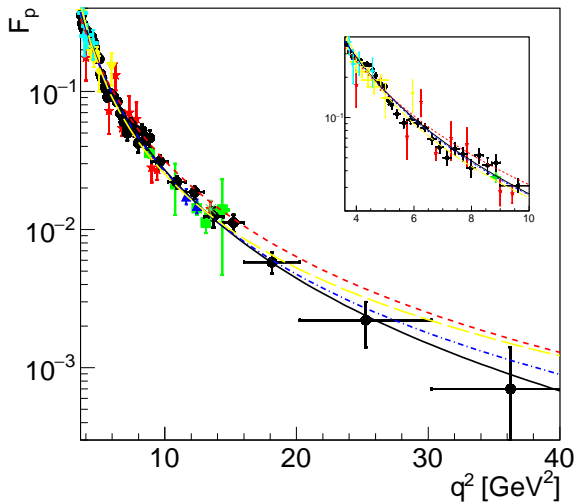}
\caption{World data on the TL proton generalized FF as a function of $q^2$, 
together with the calculation from Eq. (\ref{eq:qcd}) (blue dash-dotted line), 
Eq. (\ref{eq:eak}) (red dashed line), Eq. (\ref{eq:BdT}) (yellow long-dashed line), and Eq. \ref{eq:mpr} (black solid line). 
The world data are  
from Ref. \protect\cite{Lees:2013xe,Lees:2013uta} (black solid circles), 
Ref. \protect\cite{Ablikim:2005nn}  (red stars),
Ref. \protect\cite{Ambrogiani:1999bh} (green squares),
Ref. \protect\cite{Andreotti:2003bt} (blue triangles up),
Ref. \protect\cite{Antonelli:1994kq} (yellow triangles down ),
Ref. \protect\cite{Bisello:1983at,Bisello:1990rf} (cyan full crosses),
Ref. \protect\cite{Delcourt:1979ed} (magenta full diamonds),
and 
Ref. \protect\cite{Pedlar:2005sj}(dark green asterisk). The insert 
magnifies the near threshold region.
}
\label{Fig:WorldData}
\end{center}
\end{figure}

Based on Ref. \cite{Shirkov:1997wi}, in order 
to avoid ghost poles in $\alpha_s$ the following modification was suggested (\cite{Kuraev}): 
\be
|F_{scaling+corr}(q^2)|\ =\ 
\displaystyle\frac{\cal A}{(q^2)^2\left [ \log^2(q^2/\Lambda^2)+\pi^2 \right ] }. 
\label{eq:eak}
\ee
In this case the best fit parameters are ${\cal A}=72$ GeV$^{-4}$ 
and  $\Lambda=0.52$ GeV$^{2}$. 

In Ref. \cite{Brodsky:2007hb} a form was suggested with two poles of 
dynamical origin (induced by a dressed electromagnetic current) 
\be
|F_{T3}(q^2)|\ =\ 
\displaystyle\frac{\cal A}{(1-q^2/m_1^2) (2-q^2/m_2^2) }. 
\label{eq:BdT}
\ee
The best fit parameters are ${\cal A}=1.56$,  $m_1^2=1.5$ GeV$^{2}$ and $m_2^2=0.77 $ GeV$^{2}$. 

The TLFF data from the BABAR collaboration \cite{Lees:2013xe,Lees:2013uta} were 
obtained from the reaction 
\be
e^++e^-\ \rightarrow \bar p + p\ +\ \gamma ,
\label{eq:ISR}\\
\ee
where the photon is preferentially emitted in the entrance channel. 
These data, extending from the threshold to $q^2\approx 36$ GeV$^2$, 
are well reproduced by the  function \cite{TomasiGustafsson:2001za}:
\begin{align}
&|F_{BABAR}(q^2)|\ =\ 
\frac{\cal A}{(1+q^2/m_a^2)\left[1-q^2/0.71 \right]^2},
\nonumber
\\
&{\cal A}=7.7~\mbox{GeV}^{-4},
\ m_a^2=14.8~\mbox{GeV}^2.
\label{eq:mpr}
\end{align}
The world data are shown in Fig. \ref{Fig:WorldData} as a function of the 
transferred momentum $q^2$, and compared with the fits from 
Eqs. (\ref{eq:qcd},\ref{eq:eak},\ref{eq:BdT},\ref{eq:mpr}). 
The near-threshold region is highlighted in the insert. 

In the following, we present results with  
$F_0(p)\ \equiv\ F_{BABAR}[q^2(p)]. $
This function does not follow the expected asymptotic QCD counting rules, 
but best reproduces the BABAR data, the slope of which is steeper than $1/(q^2)^2$. 
It has to be considered as a local approximation of some more complicated function. 
We have checked that, taking any of the above background possibilities gives consistent  results  (within the errors)  although the fit has 
a smaller $\chi^2$/n.d.f. using $F_{BABAR}$. 

\begin{figure}
\begin{center}
\includegraphics[width=8.3cm]{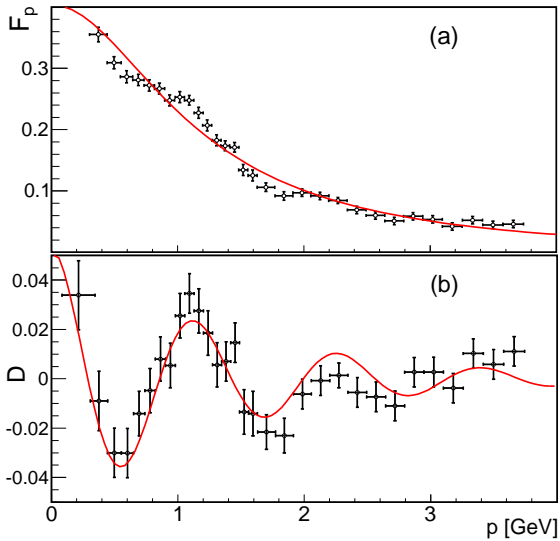}
\caption{(a): TL proton generalized FF as a function of $p$ from 
Ref. \protect\cite{Lees:2013xe,Lees:2013uta}; the line is the 
regular background 
fit with Eq. (\protect\ref{eq:mpr});  (b): data after subtraction of the fit; 
the line is the fit with Eq. (\protect\ref{eq:eqdif}).
}
\label{Fig:FitDiff}
\end{center}
\end{figure}

In Fig. \ref{Fig:FitDiff}a  the BABAR data are plotted as a function 
of $p$. The result of the fit using Eq. (\ref{eq:mpr})
is then subtracted from the 
data. This difference ${\cal D}$ (i.e., data minus $F_0(p)$) 
is shown in Fig. \ref{Fig:FitDiff}b and exhibits a damped oscillatory 
behavior, with  regularly spaced maxima and minima. 
Assuming that the first maximum is at $p$ $=$ 0, the distance between 
this, the 2nd and the 3rd maximum is $\Delta p$ $\approx$ 1.14 GeV. 
After the 3rd maximum the oscillations of the data are within the 
error bars. 

This behavior is fitted with  the 4-parameter function 
\be
F_{osc}(p)\ \equiv\ A\ \exp(- Bp)\ \cos(C p\ +\ D). 
\label{eq:eqdif}
\ee
The values  of the parameters are reported in Table \ref{tab:table}.

\begin{table}
\caption{\label{tab:table}Fit parameters  from Eq. (\protect\ref{eq:eqdif}). }
\begin{ruledtabular}
\begin{tabular}{cccccc}
\hline
    $ A \pm \Delta A $ &
    $B \pm \Delta B$ &
    $C \pm \Delta C$ &
    $D \pm \Delta D$ &
    $ \chi^2/n.d.f$   \\ 
  & 
     $[GeV]^{-1}$ &
    $[GeV]^{-1}$ &
    &
      & 
       \\ 
       \hline
     $0.05 \pm   0.01$                   
    &$ 0.7 \pm  0.2$  
    &$5.5 \pm  0.2 $     
    &$0.03 \pm 0.3$    
    &1.2  
\end{tabular}
\end{ruledtabular}
\end{table}

The relative size of the oscillating term over the regular background is 
$\sim$ 10 \%. The damping 
range of the oscillations of Fig. \ref{Fig:FitDiff}b 
is $1/B$ $\approx$ 1.4 GeV. $F_0(p)$ decreases by a factor 
$1/e$  within about 1.5 GeV. 
The relative magnitude of the oscillations to the regular 
background term $F_0(p)$ does not change much at increasing $p$, 
although increasing errors make the 
oscillations undetectable for $p>3$ GeV ($q^2 >10$ GeV$^2$). 
At asymptotically large $q^2$ values, the Phragm\'en-Lindel\"off
theorem \cite{titchmarsh1939theory} requires  that the imaginary part of 
TLFFs vanishes, which implies that rescattering disappears. 
So, although we expect a large-momentum 
suppression of the relative weight of the rescattering terms, this  
is not seen in the range $q^2< 10$ GeV$^2$ 
where error bars allow us 
to distinguish systematic from statistical oscillations in the data. 

The periodicity and the simple shape of the oscillations 
seem to exclude a random 
arrangement of maxima and minima of heterogeneous origin. Rather,  
they indicate a unique interference mechanism behind all the visible 
modulation. 
A simple oscillatory behavior in $p$ means that the waves corresponding 
to the outgoing particles originate from a small number of 
coherent interfering sources: these waves may share a common initial source but be  
rescattered along different paths or in different ways 
so to acquire different phases. 
We may speak of ``alternative rescattering pathways''. 
These must be in a small number and there must be 
discontinuity among them, otherwise we would see diffraction patterns 
instead of interference patterns. 
For example, in \cite{Kuraev:2011vq} it has been 
suggested that, during the intermediate stages of  
$\bar{p}p$ formation, charge and color are distributed in a highly 
inhomogeneous way, with discontinuity features. 

\begin{figure}
\begin{center}
\includegraphics[width=9cm]{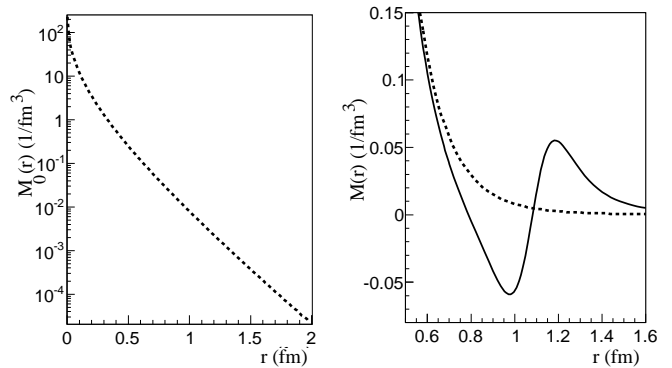}
\caption{(Left) $M_0(r)$, Eq. (\protect\ref{eq:c0}) .    
(Right): $M(r)$ (solid line) from 
Eq. (\protect\ref{eq:c1}), and $M_0(r)$ (dashed line) for comparison
(linear vertical scale).  
} 
\label{Fig:fitdoppio}
\end{center}
\end{figure}

Since we do not know the rescattering mechanism, we 
cannot identify the sources of rescattered waves, but we may gain  
some clue on their space distribution. 
Let $\vec r$ be the space variable that is conjugated to $\vec p$ via 
three-dimensional Fourier transform. 
We may identify $r$ as the distance between the centers  of 
the two forming or formed hadrons, in the frame where one is at rest. 
Let $M_0(r)$ and $M(r)$ be the Fourier transforms of the regular background 
fit and of the complete fit: 
\begin{align}
&F_0(p)\equiv\ \int d^3 \vec r\ exp(i \vec p \cdot \vec r)\ M_0(r)
\label{eq:c0}\\ 
&F(p) = F_0(p) + F_{osc}(p)\equiv\int d^3 \vec r\ exp(i \vec p \cdot \vec r) M(r).
\label{eq:c1}
\end{align}
$M_0(r)$ is shown in Fig. \ref{Fig:fitdoppio}, left panel. 
The most relevant feature is that $M_0(r)$ decreases by 7 orders of 
magnitude for $r$ ranging from 0 and to 2 fm. The decrease is 
regular and almost constant on a semilog scale. $M_0(r)$ is 
steep near the origin, too. From a mathematical point of view 
this follows from the fact that at the 
threshold of the $\bar{p}p$ channel the function 
$F_0(p)$ is a regular, continuous, and rapidly decreasing function of $q^2$. 
It can be interpreted by the fact that both $F_0(p)$ and its transform 
$M_0(r)$ are expression of that short distance quark-level dynamics  \cite{Matveev:1973uz,Brodsky:1973kr} that 
permits exclusive $\bar{p}p$ production at the condition that the final 
quarks and antiquarks are formed within a small region.  
Near threshold,  the size of this region is $\leq 0.1$ fm, much 
smaller than the standard hadron size. 

In the right panel of Fig. \ref{Fig:fitdoppio} $M(r)$ is superimposed 
to $M_0(r)$. We notice that these two functions do not differ 
for $r<0.7$ fm, and that the physical reason of the data oscillation 
must be searched for in processes taking place in the 
$r$-range 0.7-1.5 fm. 
This range is important because it includes 
the distances corresponding to the largest annihilation probability 
in the phenomenological 
$\bar{p}p$ interactions in the near-threshold region  
\cite{Bianconi:2000ap,Batty:2000vr,Friedman:2014vva}.   
At a distance of 1 fm, the relevant part of rescattering 
must involve physical or almost 
physical hadrons that annihilate into groups of 2-10 mesons. 
As discussed above, this means 
a large contribution to the imaginary 
part of the amplitude for $\gamma^*\rightarrow\bar{p}p$ from 
the Cutkosky cuts applied to all the 2-step processes like 
$\gamma^*\rightarrow n\pi\rightarrow\bar{p}p$, where $n\pi$ 
is a state composed by $n$ on-shell pions (or other mesons). 

Rescattering with a phase shift between alternative channels 
may take place via formation of 
$s$-channel poles, or via $t-$channel 
photon/meson exchanges. Both classes of processes have important 
SL continuations. 
Phenomenological $s-$channel poles lead to nontrivial phenomena due 
to their imaginary parts, but not to periodic oscillations  
(see the analysis in \cite{Brodsky:2004aa} on the  
continuation of some known SL models to the TL region).  
Exchange of $t$-channel photons leads to a Coulomb 
phase in the TL region. 
In the case of photon/meson $t$-channel exchange,  we have a large 
set of possibilities (e.g. $n\bar{n}$ formation followed by charge 
exchange). In the corresponding SL diagrams a 
virtual photon/meson is emitted 
by the nucleon before the hard vertex and reabsorbed after it. 
In the case of  virtual photons, at large (TL and SL) 
$|q^2|$ this process would not be  
more relevant than multiple photon exchange between the nucleon and 
the lepton currents \cite{Kuraev:2009aa}.  
If the exchanged bosons are mesons, 
the corresponding SL diagrams modify the 
distribution of the proton charge. The way SLFFs are affected is strongly  model dependent 
 as shown in particular 
in the neutron case in a series of works  
(see \cite{Pasquini:2007aa} and references therein). 


In summary,  a systematic modulation pattern in the TLFF measured by the BABAR collaboration in the near-threshold region has been highlighted in the range $q^2$ $<$ 10 GeV$^2$. This modulation presents 
periodical features with respect to the momentum $p$ associated 
with the relative motion of the final hadrons. It suggests 
an interference effect involving rescattering processes at moderate 
kinetic energies of the outgoing hadrons. Such processes 
take place when the centers of mass of the produced hadrons are 
separated by $\simeq$ 1 fm. For this reason 
at least a relevant part of rescattering must consist of interactions between 
phenomenological or almost phenomenological protons and antiprotons. 
These phenomenological reactions are known to have 
inelastic cross sections overcoming 1/2 of their unitarity limit. Unitarity arguments imply the presence of a large imaginary part 
of TLFF. 
The relative  errors of the data  increase with $q^2$, making us able 
to detect the modulation for $q^2$ $<$ 10 GeV$^2$, but its relative 
magnitude of about 10 \% is constant in this range, suggesting the 
interesting possibility that this modulation could be observed at larger 
$q^2$ in forthcoming more precise data. 
Precise measurements in the near threshold region are ongoing at 
BESIII (BEPCII), on the proton as well as on the neutron, bringing 
a new piece of information. The measurement of TL FFs  in a large $q^2$ range 
will be possible at PANDA (FAIR).

\end{document}